\documentclass[]{elsarticle}
\usepackage{natbib}
\usepackage{soul}
\usepackage{graphicx}

\makeatletter
\def\ps@pprintTitle{%
 \let\@oddhead\@empty
 \let\@evenhead\@empty
 \def\@oddfoot{}%
 \let\@evenfoot\@oddfoot}
\makeatother

\begin{document}

\title{Convective storms and atmospheric vertical structure in Uranus and Neptune}

\author{R. Hueso$^{1}$, T. Guillot$^{2}$ and A. S\'anchez-Lavega$^{1}$}

\address{$^{1}$F\'isica Aplicada I, Escuela de Ingenier\'ia de Bilbao, UPV/EHU, 48013 Bilbao, Spain\\
$^{2}$Universit\'e C\^ote d'Azur, Laboratoire Lagrange, OCA, CNRS UMR 7293, Nice, France}




\begin{abstract}
The Ice Giants Uranus and Neptune have hydrogen-based atmospheres with several constituents that condense in their cold  upper atmospheres. A small number of bright cloud systems observed in both planets are good candidates for moist convective storms, but their observed properties (size, temporal scales and cycles of activity) differ from moist convective storms in the Gas Giants. These clouds and storms are possibly due to methane condensation and observations also suggest deeper clouds of hydrogen sulfide (H$_2$S) at depths of a few bars. Even deeper, thermochemical models predict clouds of ammonia hydrosulfide (NH$_4$SH) and water at pressures of tens to hundreds of bars, forming extended deep weather layers. Because of hydrogen's small molecular weight and the high abundance of volatiles, their condensation imposes a strongly stabilizing vertical gradient of molecular weight larger than the equivalent one in Jupiter and Saturn. The resulting inhibition of vertical motions should lead to a moist convective regime that differs significantly from the one occurring on nitrogen-based atmospheres like  those of Earth or Titan. As a consequence, the thermal structure of the deep atmospheres of Uranus and Neptune is not well understood. Similar processes might occur at the deep water cloud of Jupiter in Saturn, but the Ice Giants offer the possibility to study these physical aspects in the upper methane cloud layer. A combination of orbital and in situ data will be required to understand convection and its role in atmospheric dynamics in the Ice Giants, and by extension, in hydrogen atmospheres including Jupiter, Saturn and giant exoplanets. 
\end{abstract}



\maketitle
\section{Introduction}

Modern explorations of possible mission scenarios to the Ice Giants Uranus and Neptune \cite{Fletcher2020, Hofstadter2019, Simon2020, Masters2014}  have motivated a revision of what we know about these planets and their planetary systems. Recent reviews explore their atmospheric dynamics \cite{Hueso2019}, mean circulation patterns \cite{Fletcher2020b}, and vertical structure \cite{Hueso2019, Atreya2020}. These atmospheric themes connect with their internal structure \cite{Helled2020} and formation mechanisms \cite{Guillot2005}. Knowledge of the properties of the Ice Giants can guide us to understand the atmospheres of the growing population of exoplanets with Neptune-size masses.

Additional insights to understand Uranus and Neptune come from the recent exploration of Jupiter and Saturn by the Juno and Cassini missions \cite{Guillot_Fletcher2020}. Both planets have zonal winds that extend a few thousand kilometers in the atmosphere \cite{Kaspi2018, Galanti2019}. For Jupiter, new questions arise on how meteorological processes distribute condensables well bellow the upper clouds and through the planet \cite{Li2017} and in both planets it is unknown how deep  latitudinal variations in the banded patterns of the planets extend. These new results and questions from the detailed exploration of Gas Giants challenge our expectations on the properties of the Ice giants Uranus and Neptune, and in particular in the distribution of condensables, the characteristics of moist convection and how these factors manifest in observable fields such as the dynamics of the cloud fields.

In hydrogen-helium atmospheres, condensables are heavier than dry air, imposing a fundamentally different dynamic regime to non-hydrogen based atmospheres \cite{Guillot1995}. In the Gas and Ice Giants, above a critical abundance of the condensing species, moist convection is {\em inhibited} by the weight of the condensables rather than favored by latent heat release \cite{Guillot1995, Leconte2017, Friedson+Gonzales2017,Guillot2019a}. This inhibition requires a sufficiently high abundance of condensables, which might be close to the water abundance in Jupiter \cite{Li2020} and might be found in the deep water cloud in Saturn, thus providing a way to explain the observations of episodic storms on this planet \cite{Li2015}. In the case of Uranus and Neptune,  methane, which condenses in the directly observable part of the atmosphere, is abundant enough to lie in that regime. Uranus and Neptune are therefore extremely interesting laboratories to understand convection in hydrogen atmospheres and its overall effect on the global atmospheric dynamics.

Recent observations of Jupiter's deep weather layer by Juno \cite{Li2017} and the Very Large Array (VLA) \cite{dePater2016, dePater2019}, and results  from Cassini following the 2010-2011 Great Storm in Saturn \cite{SanchezLavega2019, Jansen2016, Sromovsky2016} suggest that moist convection acts in both planets to desiccate NH$_3$ of the upper trospospheres below the ammonia condensation level due to details in the thermodynamics of the water ammonia system \cite{Guillot2020a, Guillot2020b}. But do other volatiles transport mechanisms occur in Uranus and Neptune cloud layers? And how observations of their cloud activity constrain the properties of moist convection in these colder atmospheres with lower internal heat sources?

Uranus and Neptune have time variable cloud systems where storms are believed to be fueled by methane condensation, but the frequency, spatial distribution and strength of these storms including their size and temporal duration are not well-known and most cloud activity observed so far is probably not convective. Methane abundance is constrained from remote sensing to values in the range of 80$\pm$20 times solar \cite{Guillot_Gautier2015,Karkoschka2011a,Sromovsky2011} (solar abundance is the proportion of a given element to hydrogen in the Sun's photosphere; the standard values of solar abundances are given in \cite{Asplund2009}), which translate in cloud bases close to 1 bar and consistent with observations of cloud systems in both planets. Models of the formation of Uranus and Neptune \cite{Helled2020} imply similar high abundances of other condensables that would form a multi-layered cloud structure with deep water clouds at pressures that can range 500-3,000 bar for water abundances in the range of 20-80 times solar \cite{Hueso2019, Atreya2020}, i.e. well above the limits to produce an inhibition of convection \cite{Guillot1995, Leconte2017, Friedson+Gonzales2017}. Both planets receive weak amounts of energy from the Sun. The highest mean daily solar irradiation received in Uranus is 3.7 Wm$^{-2}$ in its polar regions during summer, and 0.7 Wm$^{-2}$ in Neptune, also near polar latitudes \cite{Hueso2019}. The annual average solar insolation also varies strongly as a function of latitude, with Uranus receiving the highest annual average insolation in their polar regions and Neptune in the equator \cite{Moses2018}. In addition, Neptune possesses an internal energy source (it radiates $2.61\pm 0.28$ times as much energy as it absorbs from the Sun), whereas the internal flux measured at Uranus appears to be at least one order of magnitude smaller \cite{Pearl1991}. In terms of flux density, the internal heat source in Uranus is negligible, and in Neptune it is about 20 and 7 times smaller than those of Jupiter and Saturn, respectively, resulting in a much lower capacity to power internal convection.


The goals of this paper are to review the observational evidences in favor of moist convection in Uranus and Neptune, explore how moist convection may operate in the Ice Giants from a comparison with moist convection in the Gast Giants and inspect the effects that storms may produce in the vertical properties of the atmosphere. We also aim to address which measurements a mission to these planets should perform, in addition to the fundamental determination of a vertical profile of temperature and condensables abundances that could be obtained by an atmospheric probe \cite{Mousis2018}.

\section{Observations of convective activity in Uranus and Neptune} 

Here we review the observational evidences in favor of moist convective storms in Uranus and Neptune (i.e. clouds formed by vertical ascending motions powered by buoyancy differences and vertically transporting heat). These evidences come from two sources: (a) Observations of the cloud activity; (b): The inferred abundance of atmospheric methane above the tropopause for Neptune \cite{Baines1994}. Both are incomplete sources of information and show remarkable differences with what we know about convective storms in Jupiter and Saturn. 
The observations of the possible cloud activity in Uranus and Neptune either lack the spatial resolution, or the temporal sequences required to study their dynamics, or the spectral coverage to properly model the vertical cloud structure of candidate storms. 
The high  methane abundance above the tropopause was historically the main argument in favor of moist convection in Neptune. This argument does not apply to Uranus (because of a low methane abundance above the tropopause) and Jupiter and Saturn (although they are convective, they do not show high ammonia abundances above the tropopause). In addition, this high abundance in Neptune's lower stratosphere can be explained by other ideas (leakage of methane at the lower stratosphere from warm polar regions) without requiring moist convection \cite{Orton2007, Lellouch2015}. Thus, in this section we will discuss candidates to moist convective storms and not cloud systems completely proven to have a convective origin.

Observations of moist convective storms in the Gas Giants show divergent clouds on short time-scales of hours to a few days, elevated clouds above their environment and close to the tropopause, and in some cases large disturbances evolving over months. Clouds with strong divergences, as updrafts expand in the stable part of the troposphere, are a key signature of convective storms, but the observational record of such activity in Uranus and Neptune is almost inexistent because it requires frequent observation at very high-spatial resolution. In Jupiter and Saturn features with irregular cloud morphologies that show rapid changes tend to be convective, as opposed to compact ovals (closed vortices), regularly spaced spots and long filamentary wavy structures (waves). Intense convective storms large enough to be observed from Earth are frequent in Jupiter \cite{Vasavada2005} and rare in Saturn \cite{SanchezLavega2019, SanchezLavega2020}. In both planets radio emissions emitted from lightning have been observed by different spacecraft \cite{Aplin2017}, and lightning flashes have been observed in some of the most intense storms \cite{Little1999, Dyudina2013}, including Jupiter locations with no evident storms in the observed cloud field \cite{Kolmasova2018, Becker2019}, but lightning in Uranus and Neptune has only been inferred from radio signals \cite{Zarka1986, Gurnett1990, Aplin2020}.

Juno observations of Jupiter show small-scale storms at many latitudes. Storms big enough to be observed from Earth are more rare \cite{Inurrigarro2020}, and strong convective eruptions able to change the visual aspect of entire bands (like in the North Temperate Belt or in the South Equatorial Belt) \cite{SanchezLavega1996, SanchezLavega2008, SanchezLavega2017} only happen every few years \cite{Fletcher2017a}. Saturn storms occur also on different scales, with small puffy clouds in the polar regions, seasonal storms at mid-latitudes, and the much more rare Great White Spots \cite{SanchezLavega2020}. In both planets the mechanisms underlying the spatial distribution of storms, their cyclic nature in Jupiter, and their seasonal distribution in Saturn, are not known. 


Due to the difficulties to study cloud motions in images of Uranus and Neptune, the fundamental criterion for selecting candidates of convective storms is their morphology as irregular features or elongated plume-shaped clouds, with rapid temporal evolution in their size and shape and their high brightness in images at methane absorption bands indicative of high cloud tops. Besides convection, bright clouds are also observed accompanying dark anticyclones in Neptune \cite{Smith1989}. These ''companion clouds'' are suggested to be formed by similar processes to those occurring on orographic clouds on Earth. Numerical simulations indicate that methane humid air forced by the vortex circulation can rise a few kilometers and condense close to the tropopause as the vortex interacts with the environment flow \cite{Stratman2001}. These bright clouds are produced through a dynamic interaction that does not require buoyancy produced by condensation as in convective storms and are separated from the moist convective storm candidates we discuss here.

\subsection{Convective activity in Uranus}

\begin{figure}[!h]
\centering\includegraphics[width=4.5in]{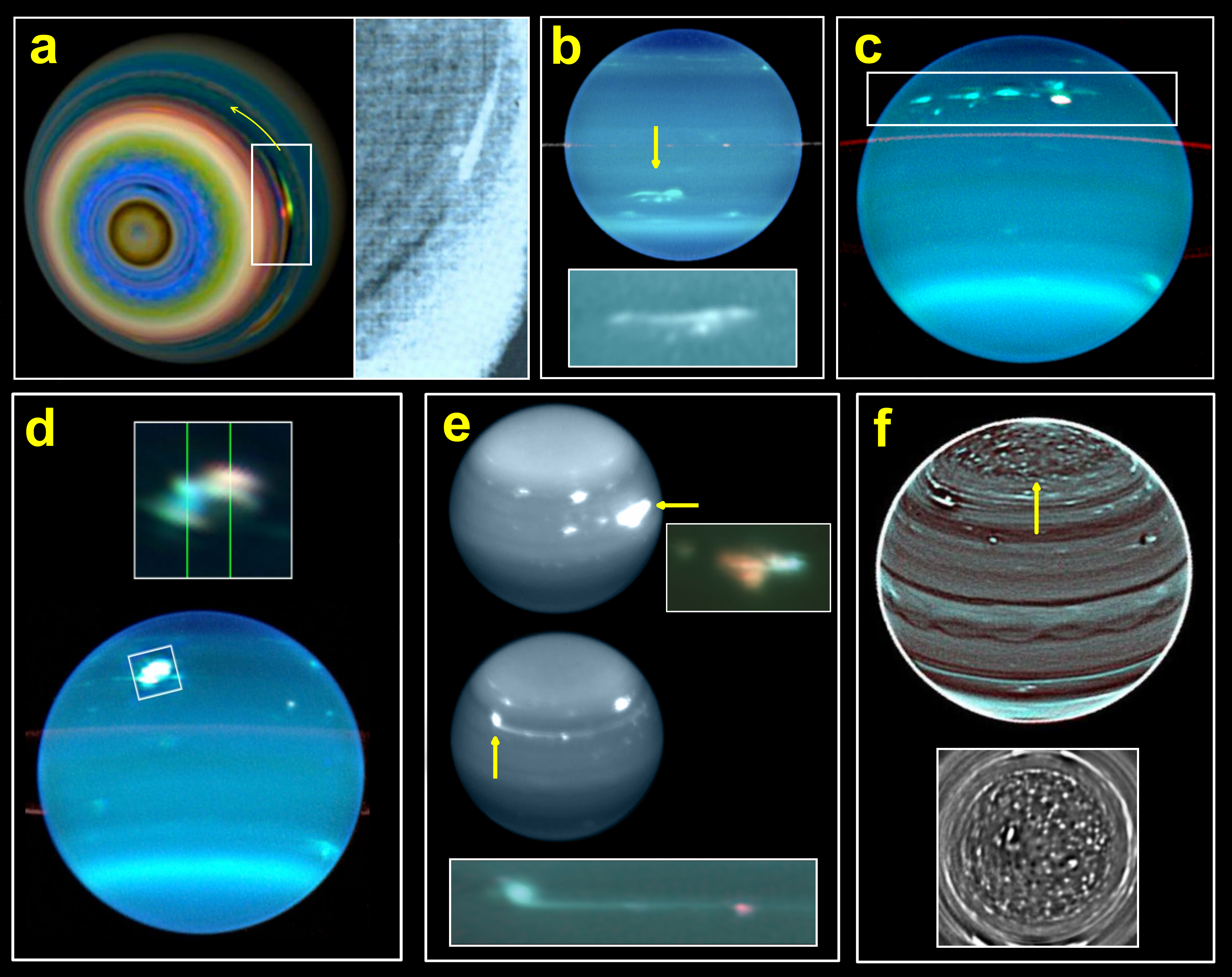}
\caption{Candidate features to moist convection in Uranus: 
(a) A plume-like feature observed by Voyager 2 in 1986 in Uranus' southern hemisphere \cite{Smith1986, Karkoschka2015}. 
The arrow signals the direction of rotation and the inset shows details from \cite{Smith1986}; 
(b) The Berg feature observed in 2007 (arrow) \cite{Sromovsky2009} with details in the inset; 
(c) The active northern band with multiple spots observed  in 2004 \cite{Sromovsky2007a, Sromovsky2007b}; 
(d) Bright spot observed in 2005 \cite{Sromovsky2007a}; 
(e) The brightest spot in Uranus as observed in 2014 \cite{dePater2015} with the elongated cloud system with insets showing details; 
(f) Cloud cluster in the North Pole observed in 2012 with details shown in the inset \cite{Sromovsky2015}. 
North is up in all the images. Panels (a-d) show color composite images based in visible to near infrared wavelengths below 1 $\mu$m with bright features in wide methane absorption bands. Panel (e) from images in band H (1.6 $\mu$m) with insets using a combination of near infrared images in bands H (blue), a CH4S  filter (1.53-1.66 $\mu$m, green) and K' (2.2 $\mu$m, red), being the latter the most sensitive to high clouds. Panel (f) shows observations in band H.}
\label{fig_uranus}
\end{figure}

The first spatially resolved images of Uranus were obtained by the Voyager 2 in 1986 and showed a planet with a homogeneous cloud cover with no bright features \cite{Smith1986}. Voyager observations of the atmospheric lapse rate and ortho to para hydrogen were used to suggest thin-layered convection with sub-saturated methane \cite{Gierasch1987}.
Those observations inspired the idea of a planet with low cloud activity, and slow variability. Figure \ref{fig_uranus} shows cloud features in Uranus that reveal that that early view of Uranus is incorrect. There are several cases of discrete bright clouds that could be caused by moist convective storms.  Some of them were short-lived and others were active for years. 

Voyager 2 only found a plume like feature at 35$^{\circ}$S whose morphology  and brightness in methane band filters were suggestive of a convective origin \cite{Smith1986} (Fig. \ref{fig_uranus}a). Later observations with ground-based telescopes and the Hubble Space Telescope (HST) captured this feature between 1994 and 2009 \cite{Hammel2005, Sromovsky2009, dePater2011} (Fig. \ref{fig_uranus}b). This feature was nicknamed the  ''Berg'', it showed oscillations in longitude and a latitude migration that eventually caused its disintegration when it reached 5$^{\circ}$S. The Berg extended $\sim 5,000-10,000$ km and had smaller clouds whose brightness changes suggested convective sources. Although most of the clouds in the Berg could be ``companion clouds'' as those occurring on Neptune's dark vortices (i.e., generated by an unobserved deep anticyclone) intense brigtening events in 2004 and 2007 suggest convective storms \cite{dePater2011}.

As the northern hemisphere turned into view in the mid-90s, a large number of discrete features were captured in a latitude band centered at 30$^{\circ}$N \cite{Karkoschka1998}. Because that latitude received the first rays of the Sun after a long winter, this activity was proposed to be seasonal convection triggered by the increasing insolation. However, observations made to date suggest that the band between 28$^{\circ}$N and 42$^{\circ}$N is an intrinsically active region in generating discrete bright spots with typical sizes of 2,000-4,000 km and sometimes developing in series \cite{Sromovsky2007a} (Fig. \ref{fig_uranus}c). The cloud tops of the brightest features (Fig. \ref{fig_uranus}d) were estimated to be at 300-500 mbar. New outbreaks of activity occurred in 2004-2006 at this latitude and were named ''Bright Complex'' \cite{Sromovsky2007b}. This high level of activity lasted until Uranus' equinox in December 2007 \cite{Sromovsky2009}. 

Another bright feature was observed in 2011 and was active for several months \cite{Sromovsky2012}. The brightest spot in Uranus so far recorded was observed in 2014 close to the equator at 15$^{\circ}$N \cite{dePater2015} (Fig. \ref{fig_uranus}e). Its reflectivity in the deep methane band K' (2.2 $\mu$m) reached 30\% of Uranus disk brightness implying that cloud tops were above the 330 mbar altitude level. The feature extended $\sim$ 17,000 km in longitude and 4,300 km in latitude and its texture was complex and formed by smaller spots of $\sim$ 2,000 km (Fig. \ref{fig_uranus}e, upper inset). A second cloud system at 32$^{\circ}$N, less bright and deeper in the atmosphere at approximately 2 bar (Fig. \ref{fig_uranus}e, lower inset), developed over months forming an elongated tail-like structure similar to large-scale convective storms in Jupiter and Saturn \cite{SanchezLavega2008, SanchezLavega2019}. Because of these characteristics, these systems are probably the best candidates for moist convective storms in Uranus. However, detailed radiative transfer analysis suggest that although the clouds reached levels close to the tropopause, their optical depth (around 1 at 0.467 nm) was not as large as it would be expected for convective features where near infinite optical depths are expected at all wavelengths. The elongated tail-like structure was found to lay around 1-2 bar not reaching the top altitudes of the main bright feature \cite{Irwin2017}.

Finally, a different type of convective activity could occur in Uranus’s North Pole, above latitude 60$^{\circ}$N, where a cluster of bright spots were observed \cite{Sromovsky2015} (Fig. \ref{fig_uranus}f). These spots have sizes of $\sim$ 500 km and are mutually separated by $\sim$ 1,000-3,000 km. The pattern strongly reminds that seen in Saturn’s North Pole \cite{Antunano2018}, so the dynamical mechanisms responsible of their formation could be similar.

From the point of view of seasonal variations, there is a global indication of enhanced cloud activity and convective candidates as the planet was reaching its equinox in 2007 and afterwards until reaching a peak in 2014. Since 2014 (corresponding to 1/12th of the Uranus year after equinox) there has not been any strong indication of convective activity, although since 2015, an occasional bright cloud system appears in the boundary of the North polar hood. 

\subsection{Convective activity in Neptune}

\begin{figure}[!h]
\centering\includegraphics[width=4.5in]{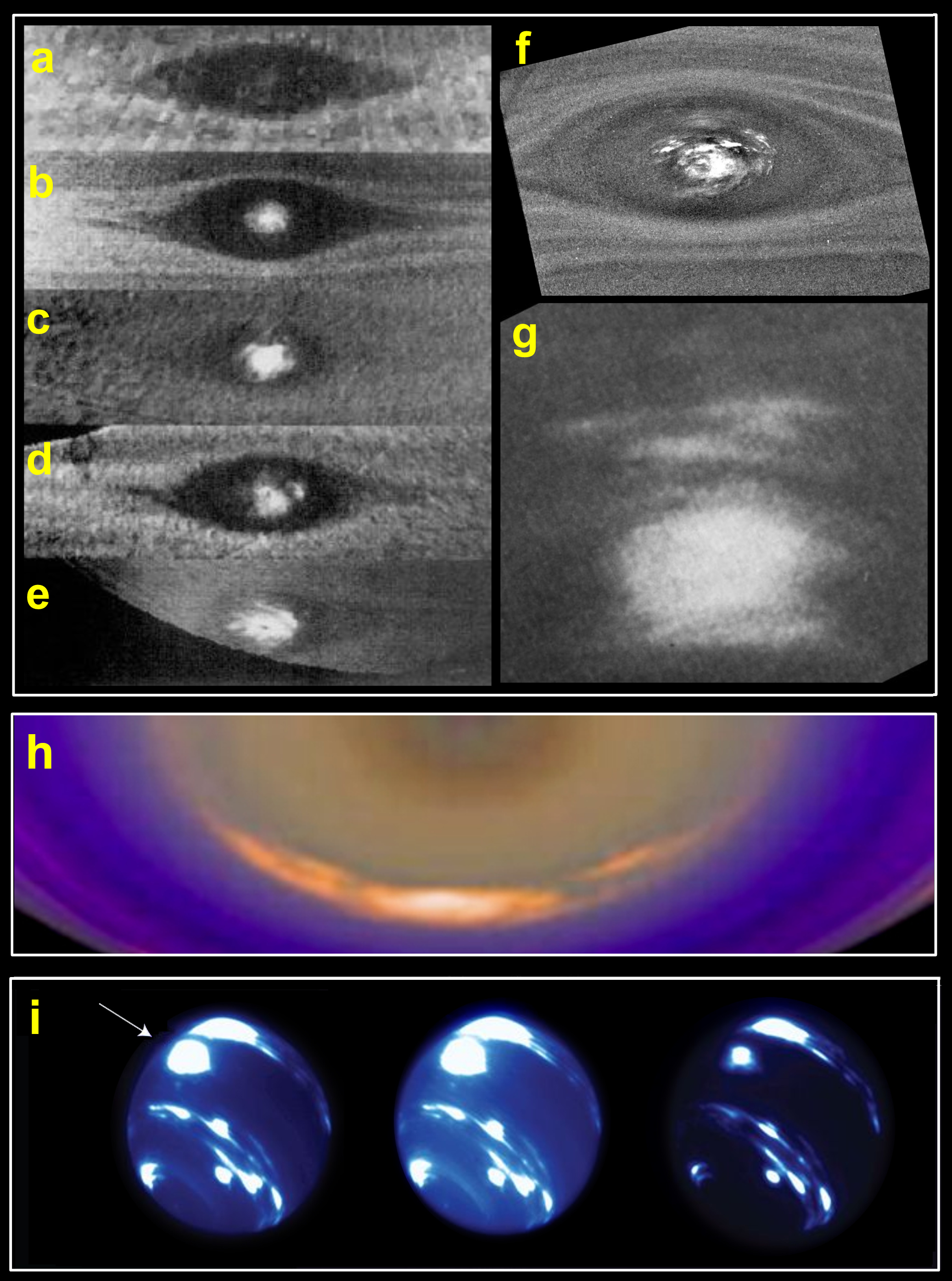}
\caption{Candidate features to moist convection in Neptune: 
(a-e) Variable activity at the center of the anticyclone DS2 in 1989 from Voyager 2 \cite{Sromovsky1993}; 
(f) Details at the center of DS2 \cite{Smith1989, Sromovsky1993}; 
(g) The ''scooter'' imaged by Voyager 2 \cite{Smith1989};
(h) Details of the SPF observed by Voyager 2 \cite{Karkoschka2011b}; 
(i) Bright equatorial cloud complex in 2017 at near infrared wavelengths \cite{Molter2019}. 
Panels (a-g) in Voyager 2 clear filters (a-b), orange (c), violet (d),  green (e), blue (f) and green (g).
Panel (h) is a false color composite with red, green and blue from observations in yellow, blue and ultraviolet light respectively.
Panel (i) show images on moderate to strong methane absorption wavelengths in  band H (1.63 $\mu$m), CH4 (1.59 $mu$m) and K' (2.12 $\mu$m).}
\label{fig_Neptune}
\end{figure}

Neptune is much more active in producing new cloud systems and cloud variations than Uranus, with large-scale changes occuring regularly \cite{Hueso2017}.
The stratosphere of Neptune has a higher abundance of CH$_4$ than the saturated abundance at the cold tropopause \cite{Baines1994}, and the stratospheric abundance of CH$_4$ in Neptune is nearly 3 orders of magnitude larger than in Uranus. Early attempts to understand these high stratospheric abundances in Neptune relied on moist convection as a mechanism able to transport methane to the stratosphere 
\cite{Lunine1989, Stoker1989}. Another possibility is that methane could leak out    
to the stratosphere through warm regions like the hot south polar region \cite{Orton2007} without the need for strong convection. However, this scenario is in conflict with circulation patterns that could explain the thermal structure observed in the planet's stratosphere \cite{Fletcher2014, Lellouch2015, Fletcher2020b}. Here we review observations of Neptune that suggest vigorous methane moist convection on discrete cloud systems.

Since the Voyager 2 flyby in 1989, Neptune has exhibited a rich variety of bright and dark spots with their related ''orographic'' cloud companions \cite{Smith1989, Sromovsky1993}. Clouds with some characteristics of convective storms were observed by Voyager 2 in at least two features. One was in the interior of the so-called DS2 anticyclone at 55$^{\circ}$S, where bright and high altitude clouds developed displaying rapidly changing morphology (Fig. \ref{fig_Neptune}a-f). The relation with the vortex dynamics is a challenging issue. This vortex was an anticyclone, but convective storms in Jupiter and Saturn that resemble the possible convection inside DS2 develop in cyclones \cite{Fletcher2017b,Inurrigarro2020, SanchezLavega2020}. Furthermore, two transient bright spots, suggestive of convective activity were also observed in 2007 in Neptune's South Pole \cite{Luszcz-Cook2016}, where the wind profile shows cyclonic vorticity.

Voyager 2 images also showed a complex system of narrow streaks stacked in latitude and forming the so-called ''scooter'' at 40$^{\circ}$S, extending about $\sim$ 3,000 km (Fig. \ref{fig_Neptune}g). 
Their clouds were deeper than its surroundings but their plume-like structure was interpreted as formed by updrafts in a region with vertical wind 
shear \cite{Sromovsky1993}. In fact, the latitude band $\sim$37-47$^{\circ}$S has also been very active in bright spot activity during the last 20 years, as recorded by a variety of telescopes \cite{Sromovsky2001}, including amateur observers \cite{Hueso2017}. These spots with sizes in the range 4,000 – 8,000 km stand out in red and near infrared wavelengths, where methane absorptions occurs, therefore implying cloud tops high in the atmosphere, but their convective origin can not be proven with the existing data.

Bright clouds are 	also found in the South Polar Feature (SPF) (Fig. \ref{fig_Neptune}g) and South Polar Waves (SPW), apparently permanent cloud systems since the Voyager 2  flyby and located at sub-polar latitudes (60-70$^{\circ}$S) \cite{Karkoschka2011b}. Both  are formed by elongated clouds that display a very stable motion that has been suggested to represent Neptune’s rotation period. The clouds have been proposed to be related to a Rossby wave, but where localized convection could occur regularly forming the small bright spots embedded on it (Fig. \ref{fig_Neptune}h).  

The brightest spot observed in Neptune was a large bright spot centered at 2$^{\circ}$N  that lasted 7 months in 2017 \cite{Molter2019} (Fig. \ref{fig_Neptune}i). It developed two equatorial bright cloud systems with the largest one showing a variable zonal extension $\sim$ 8,500-12,000 km and $\sim$ 6,000-9,000 km of meridional size and reaching the altitude level of 300-600 mbar. This cloud system was proposed as a good candidate for methane moist convection. No dark vortices that could explain the formation of these cloud systems were expected or found at the equatorial latitude of the bright spots. 

We stress that, while most of the features shown in Figures \ref{fig_uranus} and \ref{fig_Neptune} are very good convective storm candidates, we cannot make a definite statement on their convective nature. According to what we know on Jupiter and Saturn, convective storms can reach horizontal sizes $L$ in the range $L/R\sim 0.01-0.1$, where $R$ is the planetary radius. These are sizes comparable to most cloud systems described above. These convective storms should be formed by clusters of updrafts with horizontal sizes comparable to the scale height $H\sim 30-50$ km (depending on the temperature). Resolving these elements is currently impossible with ground based telescopes, HST or the future JWST, but will be in the limit of the ELT. Observations that could determine the convective nature of a particular cloud system require high-spatial resolution over multiple wavelengths from the continuum to the near IR methane absorption bands together with the time evolution of the global cloud fields. The apparent low frequency of convective candidates makes unlikely that such observations will be obtained from a combination of ground-based AO observations and HST or JWST observations, and the short flybys by Voyager 2 did not produce a dataset large enough to characterize convective candidates. Only observations from an orbiter mission seem able to provide such data \cite{Fletcher2020}.


\section{Vertical structure of the atmosphere}
The cloud systems described in the previous section are observed at levels compatible with methane condensation. In Jupiter and Saturn, thermal infrared (typically at 5 $\mu$m) sensitive to 2-4 bar and radio observations allow to glimpse in the dynamics of the clouds below. In Uranus and Neptune, the colder temperatures of the atmosphere imply that observations in the microwave to milimiter wavelengths are needed to peer below the upper clouds. Such observations using radio interferometers result in a banded structure of the planets down to 50-80 bar \cite{Tollefson2019, Molter2019b}. This deep structure is interpreted as a signature of the latitudinal distribution of condensables such as H$_2$S, and is interpreted as caused by a deep global circulation \cite{Fletcher2020b}. We here describe the expected multi-cloud layer structure of Uranus and Neptune and show caveats in the classical description of the vertical cloud structure.

The volatiles in Uranus and Neptune are methane (CH$_4$), ammonia (NH$_3$), hydrogen sulfide (H$_2$S), water (H$_2$O) and a mixed cloud of ammonium hydrosulfide (NH$_4$SH). The formation of ammonium hydrosulfide requires one molecule of NH$_3$ and one of H$_2$S and desiccates the atmosphere of the less abundant volatile. NH$_3$ is highly depleted in both planets at pressures lower than 40 bar. This is known from observations at cm wavelengths where both planets are too bright to allow for high abundances of ammonia \cite{Gulkis1978, dePater1991}. In addition, H$_2$S has been detected spectroscopically in both planets \cite{Irwin2018, Irwin2019}. This implies that the S/N ratio in their tropospheres must be higher than 1, at least down to the NH$_4$SH cloud base, which is close to the transition from the water ice to the water liquid level. Given sulfur's lower abundance in a solar-composition mixture, this means that sulphur should be enriched at least 5 times more than nitrogen compared to solar abundance. It also implies the absence of an NH$_3$ cloud. Potential atmospheric sinks for NH$_3$ and also H$_2$S complicate the interpretations that can be extracted from the absence of NH$_3$ in the upper troposphere. These sinks are a suspected liquid water ocean at pressures of 10 kilobar or higher and a ionic/superionic ocean at pressures of hundreds of kilobars \cite{Atreya2020}.

Most models of the vertical structure of Uranus and Neptune assume thermochemical equilibrium to describe the vertical location and extent of cloud layers \cite{Weidenschilling+Lewis1973,Atreya2005}. Although the complexity of cloud systems on Earth, Jupiter and Saturn \cite{Pruppacher+Klett1997, Sugiyama+2014, Li+Chen2019} shows that the thermochemical equilibrium assumption is not realistic, in the current absence of in situ or remote sensing data below the methane cloud, these models are generally used to anticipate what to expect below the methane cloud.

\begin{figure}[!h]
\centering\includegraphics[width=4.5in]{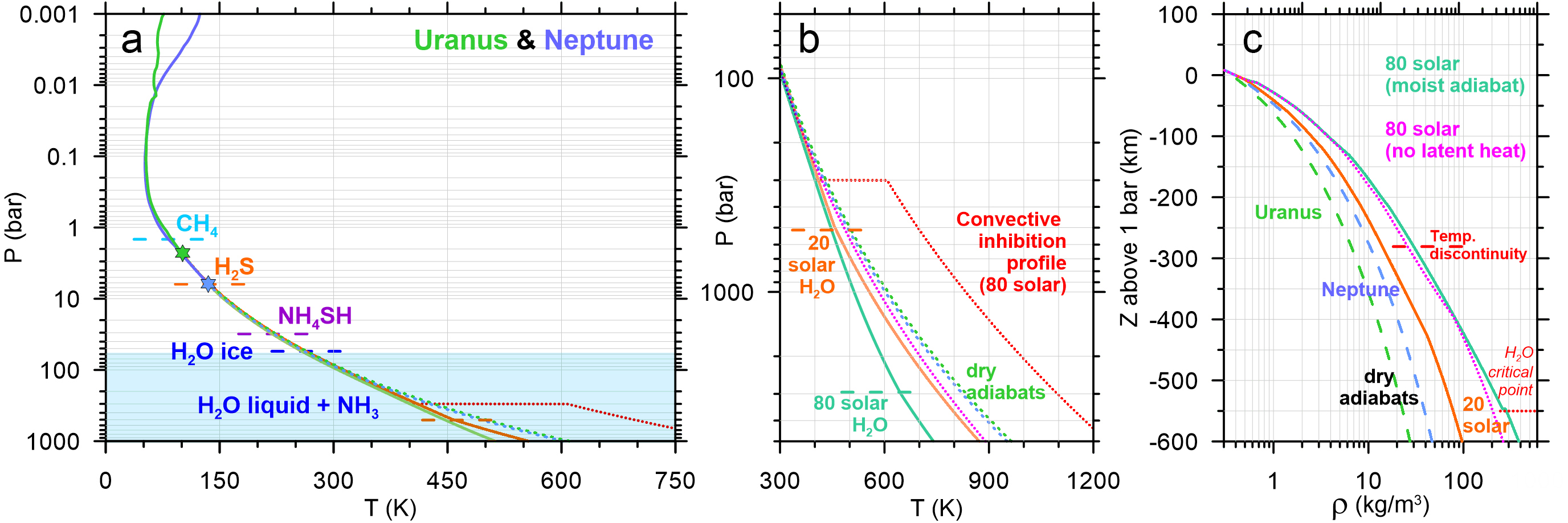}
\caption{Temperature pressure profiles measured by radio occultation in Uranus (green) and Neptune (blue) extended in the deep atmosphere following wet adiabats with different abundances of condensables. Stars in panel a show the bottom pressure layer sensed in radio occultation experiments performed by Voyager 2 in Uranus (green) and Neptune (blue). (a and b): Dashed horizontal lines represent the cloud base level for different condensates and abundances. Red profiles indicate possible thermal profiles in the case that moist convection is fully inhibited producing sharp discontinuities in temperature. These discontinuities would be found at critical abundances of methane and water but for simplicity only the one from high water abundance is shown based in \cite{Leconte2017}. c: Density profiles as a function of altitude above the 1 bar level except for the convectively inhibited profile. The kink in the 80 times solar abundance profile at 550 km is caused by the critical point of water at 647 K. Pink dotted lines in b and c show the extrapolation of Neptune's thermal profile following an adiabat without latent heat release but considering volatiles with a deep abundance of 80 times solar and relative humidities of 99\%.} 
\label{fig_thermal}
\end{figure}

\begin{figure}[!h]
\centering\includegraphics[width=4.5in]{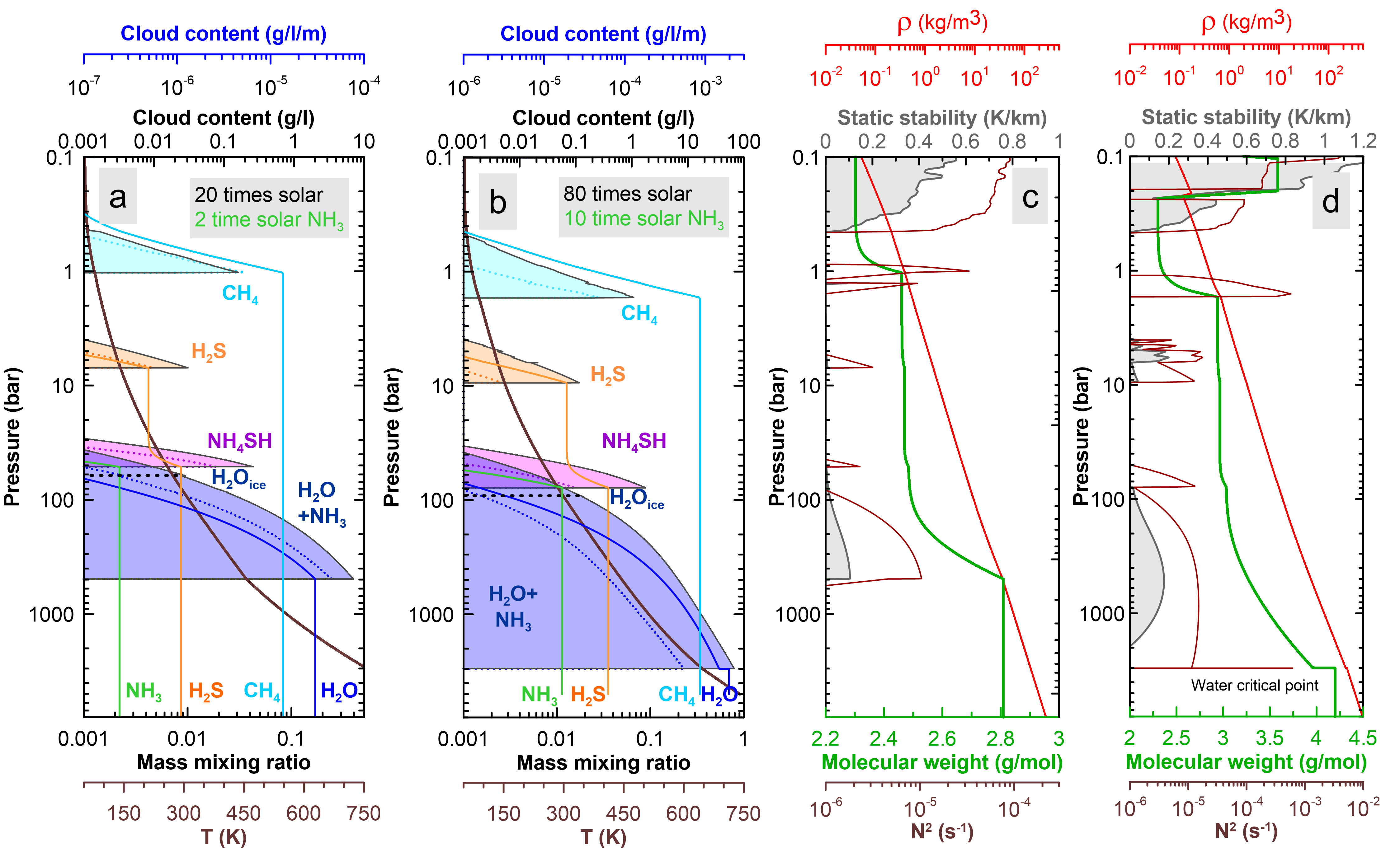}
\caption{Thermochemical models of the atmosphere of Uranus and Neptune for 20 times solar abundance of condensables (a), and 80 times solar abundances (b). Volatiles mixing ratio (colored lines, lower axis) and maximum cloud density (black lines with shaded regions, upper axis). Horizontal dashed black lines represent the transition between water ice and liquid water plus ammonia cloud. The dark red line shows the vertical temperature profile of the atmosphere (bottom axis). Maximum cloud densities can be calculated following \cite{Atreya2005} and are given in g/l (upper axis) assuming an updraft length scale equal to the atmospheric scale height and following \cite{Wong2015} (colored dotted lines) in g/l per meter (top blue axis) without assuming an explicit updraft length scale. Panels (c) and (d) show parameters related with the stability of the atmosphere for 20 and 80 times solar abundances respectively. Green lines show the vertical distribution of molecular weight (lower axis). Grey lines are the static stability of the atmosphere (upper axis). Dark red lines are Brunt-V{\"a}is{\"a}ll{\"a} frequencies including effects of the vertical gradient of molecular weight (bottom axis). Red line represents the atmospheric density profile (top axis). Note the sharp discontinuity for the model with 80 times solar abundance at the critical point of water.}
\label{fig_thermochemical}
\end{figure}

Figures \ref{fig_thermal} and \ref{fig_thermochemical} show the thermal, density and volatile vertical structure of Uranus and Neptune.
Thermal profiles obtained from the Voyager 2 radio occultation experiments \cite{Lindal1987, Lindal1992} ended at pressures of 2.3 bar in Uranus and 6.3 bar in Neptune (highlighted with stars in figure \ref{fig_thermal}a). These levels correspond to the regions where CH$_4$ and H$_2$S condense, and future atmospheric occultation measurements that could be obtained by an orbiter could sense the thermal structure of the atmosphere over different latitudes sampling both cloud condensing regions. Below those pressures we examine the planets' deep tropospheres, considering abundances of 20 and 80 times solar for all volatiles except for ammonia, which is assumed to be less abundant than H$_2$S, to be consistent with the spectroscopic detections of H$_2$S vapor in the tropospheres of both planets.

Other physical processes, including the  inhibition of moist convection at high abundances of methane \cite{Guillot1995, Leconte2017, Friedson+Gonzales2017} and water or layered convection \cite{Gierasch1987} can affect the vertical structure producing small and large thermal and mean molecular weight discontinuities in the vertical thermal profile. Figure \ref{fig_thermal}b sketches possible effects caused by the vertical distribution of water and is based in \cite{Leconte2017}. It must be noted that these large thermal discontinuities can be produced in combination with a discontinuity in the mean molecular weight of the atmosphere and the volatiles concentration without neccesarily resulting in a large discontinuity in density. Such discontinuities would be readily found by a descending atmospheric probe, but could be different at different locations of the planet requiring a characterization of the descending region from remote sensing observations by an orbiter. Figure \ref{fig_thermal}c examines the atmospheric density profile for a range of possible deep abundances of condensables (dry atmosphere, 20 and 80 times solar). The density profile depends on the thermal structure, the local gravity and local winds (we here considered the equator and Voyager 2 winds constant in depth). Because of water's potential high abundance in the deep tropospheres of the Ice Giants, it probably dominates the overall mean molecular weight and controls the density profile.

Figure \ref{fig_thermochemical} shows the vertical distribution of volatiles and clouds compatible with the thermal structure described above. This figure does not take into account super adiabatic effects or thermal discontinuities associated to the inhibition of moist convection and can be considered as highly ideal. The densities associated to the cloud content should be considered as illustrative, as the real properties of a cloud depend on microphysical processes including the density of condensation nucleii and precipitation, which are neglected in these simple models. Properties of real clouds depend also on the history of the cloud formation and the vertical motions through the clouds. Densities in Figure \ref{fig_thermal}c do not include contributions from the clouds.

In both Uranus and Neptune the base of the water cloud lies at pressures of several hundred to a few thousand bars and their study would require the use of non-ideal gas laws. Most of the water cloud particles would be in liquid phase with the potential to dissolve ammonia, desiccating the upper atmosphere from this volatile. For the models with higher enhancement of volatiles (80 times solar) the water cloud base deepens until reaching temperatures close to the critical point of water at 647 K at pressures of $\sim3,000$ bar at a depth of about 550 km bellow the 1 bar level. The right panels in this figure show different variables related with the stability of the atmosphere. The most relevant of this is the vertical gradient of the molecular weight of the atmosphere. As a result of the increase in methane abundance, the mean molecular weight in the upper troposphere varies from 2.3 g mol$^{-1}$ in the tropopause to 2.5-2.9 g mol$^{-1}$ at the methane cloud base for 20-80 times solar abundances, respectively. The mean molecular weight continues to increase in the lower atmosphere and can reach 2.8-4.2 g mol$^{-1}$ at the water condensation level for the range of abundances here explored. For 80 times solar abundances the water cloud base occurs at the water critical level, an effect expected in many exoplanets with Neptune-like masses \cite{Mousis2020}.

High values of the deep water abundance result in a combination of high molecular weight (Figure \ref{fig_thermochemical}b-d) and lower deep temperatures for profiles where moist convection is operating (Figure \ref{fig_thermal}b). The high molecular weight determines a high density at depth expected for high water abundances suggesting that gravity measurements could be used to uncover the deep abundance of volatiles.

\section{Moist Convection in the Ice Giants}
Because of the higher stratospheric methane on Neptune than in Uranus \cite{Baines1994}, moist convective models have been traditionally developed for Neptune without additional work to address Uranus convective systems.  Different authors \cite{Lunine1989, Stoker1989} have used one-dimensional models of moist convection of methane, or simple estimations of the maximum Convective Available Potential Energy (CAPE) \cite{Molter2019} that predicted powerful updrafts capable to reach vertical velocities of $\sim 200-250$ ms$^{-1}$. These updrafts would be able to ascend to the tropopause and transport methane to the lower stratosphere. However these storms need strong perturbations of the environment to be initiated (initial vertical velocities $> 40$ ms$^{-1}$) to counteract the static stability of the atmosphere, very favourable environments with high humidity to counter-act the weight of methane gas in the saturated updraft, and very efficient rain-out . These pre-Voyager predictions of such powerful methane moist convection storms have not been verified observationally.

An additional complication is that Uranus and Neptune abundances of methane imply an inhibition of moist convection at pressure levels just below those directly observed \cite{Guillot1995}. The moist convection inhibition criteria also applies in the pressence of double-diffusive convection \cite{Leconte2017, Friedson+Gonzales2017} and it is yet unclear how heat is transported in such a situation.


In order to get a simplified grasp of the inhibition of convection we will calculate a few numbers that will help us to show the regions of the atmosphere where convective inhibition plays a role and we extend this discussion to the deep clouds that determine the vertical structure of the atmosphere. We closely follow \cite{Guillot2019a} in this discussion, and ignore the negative buoyancy effect of the weight of condensates, assuming precipitation is efficient. There are a number of quantities important to examine the potential to drive moist convection. One of them is the capacity to produce increases in temperature by the release of latent heat. The increase of temperature by latent heat from a condensable $i$ reaching its vapor saturation, $\Delta T_{Li}$, can be calculated as:

\begin{equation}
\Delta T_{L_i}=\frac{q_{vi} L_{Vi}}{C_p}, 
\end{equation}
where $q_{vi}$ is the maximum mass mixing ratio of the condensate $i$, $L_{Vi}$ is the latent heat per unit mass
and $C_p$ is the atmospheric heat capacity per mass. 

A second important quantity is the temperature increase required to compensate for the weight of the volatiles in a wet parcel, $\Delta T_{\mu i}$. This quantity can be calculated as:
\begin{equation}
\Delta T_{\mu i}=- \left[ ln(1- \varpi q_v) \right] T,
\end{equation}
where $T$ is temperature and $\varpi=(M_{vi}-M_a)/M_{vi}$ is the reduced mean molar mass difference calculated from the molar masses of the vapor $i$, $M_{vi}$, and the dry air, $M_a$. 

The third quantity is the moist convection inhibition factor $\xi_i$ \cite{Guillot1995, Leconte2017, Guillot2019a}, which predicts that moist convection for condensable $i$ is not possible when $\xi_i>1$. This inhibition factor is defined as:

\begin{equation}
\xi_i=\frac{\varpi M_{vi} L_{vi} }{ RT}q_{vi},
\end{equation}
where $R$ is the ideal gas constant. 

For models shown in Figure \ref{fig_thermochemical} these quantities are given in Table 1.  For ammonia we assume a slightly different atmosphere with equal ammonia and hydrogen sulfide resulting in an ammonia cloud with no hydrogen sulfide cloud. For those condensables where $\xi_i>1$ at the cloud base there is always a level above where $\xi_i=1$. This level constitutes the deepest level where moist convection can be powered for each condensable and is also given in Table 1 as $P_{c}$. This can be compared with the cloud base $P_b$ at the deepest layer where saturation is reached. So, although moist convection is inhibited at the base of different clouds, convection is possible in the upper atmosphere in a range of altitudes, possibly  resulting in a complex thermal and compositional structure of the atmosphere. Adding the mass-loading of precipitation would complicate further the problem and numerical models will have to be designed to answer these questions. In addition, dry convection without latent heat release could be activated by differences in condensable abundances below their saturation values with moist air sinking and dry air ascending, an effect that could also be observable in the methane condensation layer at low optical depth.

\begin{table}[]
\caption{Convective parameters for different condensables for models of Uranus and Neptune for 20 times solar abundances (1; left columns) and 80 times solar abundances (2; right columns). (*) Ammonia clouds are explored considering a 20 times solar abundance of all condensates (1) or 80 times solar abundance (2).}
\label{table_convection}
\begin{tabular}{l | c  c  c  c  c | c  c  c  c  c}
\hline
Vapor  & $\Delta T_{L1}$  & $\Delta T_{\mu1}$ & $\xi_1$  & $P_{c1}$  & $P_{b1}$   &  $\Delta T_{L2}$  &  $\Delta T_ {\mu2}$  & $\xi_2$  & $P_{c2}$  &  $P_{b2}$ \\
           &   (K)                   &  (K)                    &              & (bar)        &   (bar)            &   (K)                   &     (K)                     &              & (bar)       & (bar)\\
\hline
CH$_4$          & 4.0     & 5.8  & 1.05  &   1.1    & 1.1   & 18.8       &  33.8    &   3.1    &  1.5     &    1.8  \\
H$_2$S          & 0.3     & 0.6  & 0.08  &    7.0   &  7.0  & 1.0         &  25       &   0.2    &  9.4     &    9.4   \\ 
NH$_3 (*)$    & 3.0     & 3.0  & 0.3    &   20     &   20   & 14.5       &  13.9    &   1.1     & 34     &   37       \\
H$_2$O         & 34      & 73   & 1.5    &   334   &   490 & 112        &  454     &   2.7     & 415    & 3000  \\
\hline
\end{tabular}
\end{table}

\section{A complex weather layer}

Given current unknowns in the abundances of volatiles in Uranus and Neptune, their impact in the vertical structure of the atmosphere, and the complexities of moist convection previously examined, a large range of possible structures could form below the upper methane clouds in Uranus and Neptune. Which measurements would provide the most data to solve open questions? We turn to Jupiter and its better known atmosphere and later examine the measurements needed for Uranus and Neptune.

\subsection{Lessons from the Galileo Probe and Juno}


In 1995 the Galileo probe entered Jupiter's atmosphere obtaining measurements of its composition and vertical structure down to 22 bars \cite{Young2003}. Measurements of ammonia and water abundances increased in depth but were separated from saturated profiles. While a deep value of ammonia was found, water was still highly subsolar and was still rising at the 22 bar level \cite{Wong2004}. Explanations of these measurements required the probe to have descended in a dry location in the planet caused by local meteorology \cite{Showman1998}. The Galileo probe did not find the deep value of oxygen, which, given the paramount importance of water for planet formation, also motivated the Juno mission.

Recent measurements of the ammonia latitudinal and vertical distribution obtained by the Juno MWR instrument resulted in another puzzle. Ammonia is not uniformly distributed in Jupiter's atmosphere below the clouds, but depleted  to around 30 bar at all latitudes except at the equator \cite{Li2017}. A viable model to explain this global desiccation is that water-powered moist convective storms could trap ammonia in hailstones that would precipitate and deplete the upper atmosphere of ammonia \cite{Guillot2020a, Guillot2020b}, making water-powered moist convective storms an essential piece of the jovian meteorology. This mechanism is in agreement with the ammonia and lower clouds aerosol depletion observed in Saturn after the development of the large Great White Storm of  2010-2011 \cite{SanchezLavega2019, Jansen2016, Sromovsky2016} and could be a general drying mechanism affecting also the atmospheres of Uranus and Neptune in case moist convection from a lower cloud could reach the condensation region of a volatile of the upper atmosphere.  The recent determination of water abundance in Jupiter ($2.7^{+2.4}_{-1.7}$ times solar, representative of the equatorial region and possibly of the deep atmosphere) \cite{Li2020} seems to exclude inhibition of water moist convection in Jupiter, which requires a water abundance of 9.9 times solar \cite{Leconte2017}.

The atmospheres of Uranus and Neptune have the potential to be more complex than those of Jupiter and Saturn due to the larger values of condensable abundances, higher number of cloud layers, and the potential transition of liquid water to vapor at the water critical point for water abundances close to 80 times solar. The higher abundance of H$_2$S than NH$_3$ in the upper troposphere requires ammonia depletion mechanisms that could be similar or different to those proposed in Jupiter and could include dissolution of ammonia in the liquid water cloud layer or in the interior below the hydrogen-helium atmosphere.

An exploration of the atmospheres or Uranus or Neptune by an atmospheric probe \cite{Mousis2018, Atreya2020} would be limited to the first few tens of bars. Radio interferometric measurements performed by ALMA show the structure of Neptune down to 80 bar at low spatial resolution, and we have seen that the deep abundance of water plays a critical aspect in the overall density profile that could be investigated from a gravity experiment. Measurements from an in situ probe would require a global characterization by an orbiter \cite{Fletcher2020} sensitive to the deep structure of the atmosphere through a combination of gravity measurements and microwave radiation. Such an orbiter could also obtain detailed observations of the methane clouds and convective storms and could determine through occultation experiments thermal profiles at several locations of the planet to depths of a few bars. In Jupiter, the combination of results from the Galileo Probe, the Juno mission and the detailed observations of jovian meteorology at the upper ammonia clouds show that such a combination of global and detailed local measurements is needed to understand its atmosphere. The alternative to send a secondary or multiple probes at other planet locations would also highly enhance the capability to study these questions \cite{Sayanagi2020}, but would also benefit from a characterization from orbit studying also changes in the upper atmosphere.

\subsection{Non-homogeneous weather layers}

The standard picture of Uranus and Neptune used during decades assumes well-defined cloud decks and a temperature profile that follows a moist adiabat accounting for the condensation of the different species. Meridional circulation may affect the abundances of condensables and the thermal properties, producing contrasts in the band pattern of the planet at a variety of depths \cite{Fletcher2020b, Tollefson2019, Molter2019b}. An alternative and more complex picture can be build by assuming non-homogeneous weather layers where cold dry air could coexist with warm humid air at the same level, and where convective storms could shape the characteristics of the atmosphere  \cite{Guillot2019a}. The vertical structure of the atmosphere could have a combination of the following characteristics. For each one we briefly suggest measurements that could help to advance in each of these cases.

\begin{itemize}

\item     \textbf{Variable temperature profile from moist adiabatic to superadiabatic.} An orbiter capable to obtain occultation measurements could investigate the thermal profile on different locations at least to the methane cloud base.

\item     \textbf{No ubiquitous methane cloud deck but regular formation of methane clouds including possible convective storms.} These storms could combine strong updrafts powered by latent heat release and downdrafts powered by the weight of the condensates and evaporative cooling. Observations obtained from an orbiter could determine which of the active cloud systems are actually convective storms and the characteristics of such storms. The risk here is that we do not know the frequency and time scales of convective events in Uranus or Neptune.

\item    \textbf{Deep H$_2$S stable clouds with little impact in the thermal structure.} Depending on the abundance of CH$_4$ and H$_2$S, thermal profiles from occultation measurements could investigate this question. Observations in the thermal infrared from thermal emissions emitted from the lower atmosphere above the NH$_4$SH cloud level could also reveal the dynamics of this cloud layer.

\item   \textbf{Complex dynamics at the NH$_4$SH cloud layer.} This cloud layer could be complex due to the combination of a large heat release (comparable to the water latent heat release) and the high molecular weight of its components. This cloud region can be investigated by thermal emissions in the milimeter and microwave regime and a Juno-like MWR instrument could investigate the physical processes in this cloud layer through the planet. Current interferometer telescopes like VLA and ALMA are currently able to latitudinally map Uranus \cite{Molter2019b} and Neptune \cite{Tollefson2019} down to 50-80 bar.

\item   \textbf{Strong vertical transport at the water condensation layers.} The dynamics here could be an extreme version of the dynamics produced at the methane condensation region but these deep layers are hidden to observations using radiation or measurements from an atmospheric probe. Gravity measurements by an orbiter could investigate the deep abundance of water and a detailed model-observations comparison may be required to interpret gravity data.

\item   \textbf{Vertical motions driven by compositional differences and not latent heat release.} In atmospheres with large horizontal compositional gradientes vertical motions could be driven by compositional differences and not latent heat release. Updrafts of dry air in a more wet environment could be possible at any of the condensation regions. Detailed observations of these processes at the upper CH$_4$ layer could be obtained by observations in the visible and near infrared.

\item   \textbf{Ammonia powered storms in exo-Neptunes.} Cold exo-Neptunes are likely very common planets \cite{Suzuki2016}, and they may be very diverse in terms of the abundances of volatiles.  In those planets with higher tropospheric abundances of NH$_3$ than H$_2$S ammonia moist convective storms could have similar strength to methane storms in Uranus and Neptune. The lack of observed NH$_3$ in Uranus and Neptune upper tropospheres implies that the best candidates to study ammonia-driven moist convective storms are Jupiter and Saturn, where ammonia storms seem to play a minor role compared with water powered storms.
\end{itemize}
\vspace*{-5pt}

\section{Conclusions}

The mechanisms responsible for the structure, dynamics and energy balance of the atmospheres of Gas and Ice Giants are complex and poorly known. In particular, the role of moist convection and precipitation, its importance to determine the vertical structure of temperature, condensables and density, and the interplay of moist convection with the large-scale circulation are yet to be understood. Adding to this complexity, multiple cloud layers and sources of storms are present in these planets, most of them too deep to be probed directly. This is for example the case of water, thought to be fueling most of Jupiter’s and Saturn’s storms, but whose abundance in both planets is poorly constrained because of its condensation at relatively high pressures and high optical depths. Owing to their larger distance to the Sun, Uranus and Neptune possess colder atmospheres with abundant 
methane cloud activity that could be interpreted as convective, but the existing data does not allow to determine which of the possible storm candidates observed are actually moist convective events. This methane condensation region is at a relatively low optical depth, and can be probed relatively easily. We thus believe that Uranus and Neptune are laboratories to understand the underlying physics of atmospheric dynamics in hydrogen atmospheres. The hidden cloud layers below the upper methane clouds have a key importance in the global vertical structure of the planets. Understanding these cloud layers has applications to understand and constrain the atmospheres, interior structure and evolution of not only Uranus and Neptune, but also Ice Giants and Gas Giants in general. 

Uranus and Neptune remain the only planets in the Solar System that have not been visited by an orbiting spacecraft. A mission to at least one of them would bring an essential piece of the puzzle to complete the inventory of our Solar System, understand the structure and atmosphere dynamics of Ice Giants and Gas Giants in the Universe. Such a mission should include an orbiter to fully map and characterize the planet’s atmosphere and its time-variability \cite{Fletcher2020}, and an atmospheric probe \cite{Mousis2018}, or multiprobe system \cite{Sayanagi2020}, to measure with high precision the thermal structure and volatile abundance profiles at a well-defined location, or set of locations. An instrumentation suite similar to the one carried by Juno would be the most efficient to provide spatial context to the measurements obtained by a descending probe. The combination of this information would provide unique data to understand planetary atmospheres. 


\vskip6pt

\enlargethispage{20pt}


\textbf{Acknowledgements}. RH and ASL were supported by the Spanish MINECO project AYA2015-65041-P and PID2019-109467GB-I00 (MINECO/FEDER, UE) and Grupos Gobierno Vasco IT1366-19. We are thankful to the Royal Astronomical Society for hosting the Ice Giant Systems workshop on January 2020 and to Leigh N. Fletcher for organizing this event. 



\end{document}